\documentclass[aps,prb,twocolumn,superscriptaddress,floatfix,showpacs]{revtex4}

\usepackage{graphicx}

\begin{document}

\title{Exact diagonalization analysis of the Anderson-Hubbard model\\
and comparison to real-space self-consistent Hartree-Fock solutions}

\author{X. Chen}
\affiliation{Department of Physics, Queen's University, Kingston ON K7L 3N6 Canada}
\author{P. W. Leung}
\affiliation{Department of Physics, Hong Kong University of Science and Technology, Clear Water Bay, Hong Kong}
\author{R. J. Gooding}
\affiliation{Department of Physics, Queen's University, Kingston ON K7L 3N6 Canada}

\date{\today}

\begin{abstract}
We have obtained the exact ground state wave functions of the Anderson-Hubbard model 
for different electron fillings on a 4$\times4$ lattice with periodic boundary conditions.
When compared to the uncorrelated ground states (Hubbard interaction set to zero) 
we have found evidence of very effective screening, producing smaller charge inhomogeneities 
due to the Hubbard interaction, particularly at 1/2 filling, and have successfully modelled 
these local charge densities with
non-interacting electrons that experience a static screening of the impurity potentials. 
Further, we have compared such wave functions to self-consistent real-space unrestricted 
Hartree-Fock solutions and have found that these approximate ground state wave functions are 
very successful at reproducing the local charge densities, and may indicate the role 
of dipolar backflow in producing a novel metallic state in two dimensions.
\end{abstract}


\maketitle

Many transition metal oxides display phenomena that are believed to be associated with
both strong electronic correlations and disorder. Various metal-to-nonmetal
transitions,\cite{imadaRMP} and some properties of the underdoped high-T$_c$ cuprate 
superconductors\cite{BobBrmp,goodingPRB97,laiPRB98} are examples of such physics. 
Presently the study of such systems is a very active field of research in condensed matter
physics.

Theoretically, the simplest model that hopefully represents some of the key physics
of such systems is the so-called Anderson-Hubbard Hamiltonian. The Anderson 
model\cite{PWA} is given by
\begin{equation}
\label{eq:Ham_AndHub}
{\hat H_A}~=~\sum_{i,\sigma}~V_i~{\hat n}_{i,\sigma}~-~t~\sum_{\langle i,j \rangle,
\sigma}~\Big({\hat c}^\dagger_{i,\sigma}~{\hat c}_{j,\sigma}~+~h.c.\Big)
\end{equation}
where $i,j=1\dots N$ denote the sites of the lattice, $\langle i,j \rangle$ implies
that $i$ and $j$ are near neighbours, ${\hat c}_{i,\sigma}$ (${\hat n}_{i,\sigma}$) 
is the destruction (number) operator for an electron at site $i$ with spin $\sigma$,
and the hopping energy is denoted by $t$. The on-site energy at site $i$ is given by 
$V_i$, and in this report we have examined a 50/50 binary alloy model, 
where $V_i$ is set equal to $W/2$ (for an $A$ site) or $-W/2$ (for a $B$ site). 
The particular complexion of disorder that is used in this
study can be seen in Fig.~2 -- the filled circles represent the A sites, while the
open circles represent the B sites. 
The electron interactions that are included are represented by the Hubbard term, given by
\begin{equation}
\label{eq:Ham_Hub}
{\hat H_H}~=~U~\sum_i~{\hat n}_{i,\uparrow}{\hat n}_{i,\downarrow}
\end{equation}
The Anderson-Hubbard model is formed from the sum of ${\hat H_A}$ and ${\hat H_H}$.

There are two dimensionless energy scales in this problem: $W/t$ and $U/t$. Due 
to the challenge of the computational task (see below), in this report we present results 
for $U/t=8$ only -- such an interaction energy is equal to the bandwidth of the 
non-interacting ordered systems, that being $8t$. We have examined the weak to strong disorder 
regimes using $W/t~=~4,6,7,8,9,10$ and 12, and have completed calculations for fixed 
electronic densities of 1/4, 3/8 and 1/2 filling. 

In our numerical work we have studied a binary alloy for a 4$\times4$ lattice with periodic boundary conditions,
and have included an on-site Hubbard repulsion energy equal to the bandwidth of the ordered,
non-interacting problem. Due to the lack of translational periodicity, one must determine
all of the 16-choose-8 squared (for 1/2 filling), or roughly 166 million probability amplitudes,
and to date there are no exact wave functions available for such a large spatially disordered system.
(While work on ordered systems\cite{EDordered} has indeed studied larger Hilbert spaces,
previous work\cite{previousED} on this and related (spinless fermion) spatially inhomogeneous
Hamiltonians have not surpassed Hilbert spaces of dimension $\sim3\times 10^{5}$, less than
1/500$^{th}$ the size that we have used.) The role of exact diagonalization computational work 
in the elucidation of the physics of strongly correlated electronic systems is well known --
it provides numerically exact solutions that are free from approximations.
The only drawback is that relatively small systems must be examined. Larger systems can indeed
be studied using quantum Monte Carlo simulations, and have been completed in both two\cite{2dqmc}
and three\cite{3dqmc} dimensions. Our data is complementary to such work.
Therefore, these results can now be used as benchmarks gauging the success of proposed theories, 
such as Hartree-Fock, which is discussed below, or the popularly used dynamical mean field theory 
(DMFT).\cite{dmft} Some analogous benchmarking comparisons of DMFT for ordered systems have 
recently been completed.\cite{pothoff}

For this cluster one can exactly diagonalize the Hamiltonian matrix and find the
ground state wave function using the Lanczos algorithm. Besides the energy of
these states, we have found the local charge densities as well as one
characterization of the magnetic properties given by ${\bf S}_i\cdot{\bf S}_j$ for
$i$ and $j$ being nearest neighbours.

\begin{table}
\begin{tabular}{llll}
(a)      \\[4pt]
1.755 & $\,$ 0.140 & $\,$ 0.284 & $\,$ 1.825 \\[2pt]
0.262 & $\,$ 0.189 & $\,$ 1.778 & $\,$ 1.752 \\[2pt]
0.177 & $\,$ 1.714 & $\,$ 0.173 & $\,$ 0.261 \\[2pt]
1.842 & $\,$ 0.251 & $\,$ 1.771 & $\,$ 1.827 \\[10pt]
(b)      \\[4pt]
1.082 & $\,$ 0.941 & $\,$ 0.906 & $\,$ 1.057 \\[2pt]
0.919 & $\,$ 0.925 & $\,$ 1.091 & $\,$ 1.080 \\[2pt]
0.925 & $\,$ 1.105 & $\,$ 0.913 & $\,$ 0.918 \\[2pt]
1.075 & $\,$ 0.910 & $\,$ 1.094 & $\,$ 1.060 \\[10pt]
(c)      \\[4pt]
1.064 & $\,$ 0.959 & $\,$ 0.919 & $\,$ 1.039 \\[2pt]
0.937 & $\,$ 0.941 & $\,$ 1.079 & $\,$ 1.063 \\[2pt]
0.941 & $\,$ 1.095 & $\,$ 0.925 & $\,$ 0.935 \\[2pt]
1.061 & $\,$ 0.922 & $\,$ 1.079 & $\,$ 1.041 \\
\\
\end{tabular}
\caption{\label{table:chargesVeq4halffilling} In (a) the exact local charge densities are listed for
the $U/t=0$ non-interacting ground state at 1/2 filling for $W/t=4$; (b) shows analogous
data, but now for the exact interacting systems with $U/t=8$; (c) shows the local charge
densities for the interacting problem, now calculated within the Hartree-Fock approximation.}
\end{table}

The disorder potential without the Hubbard energy is sufficient to produce a strongly
inhomogeneous system, even for moderately small disorder. This is demonstrated in
part (a) of TABLE~\ref{table:chargesVeq4halffilling} in which we have listed the local charge densities
for $W/t=4$ and $U=0$. The A sites have charge densities from 0.14 to 0.28, whereas the
B sites have local charge densities from 1.71 to 1.84. How effective is the Hubbard
interaction energy in screening the disorder potential? 
For $U/t=8$ the answer from exact diagonalization is found in part (b) of 
TABLE~\ref{table:chargesVeq4halffilling}. With the Hubbard interaction the local charge densities 
for both A and B sites now range from only 0.91 to 1.11, and inspection
of the data for the non-interacting and interacting charge densities makes evident the
effectiveness of the Hubbard energy in producing a much more uniform charge density.

However, it is to be emphasized that when one is away from 1/2 filling the homogenization of
the local charge densities is far from complete. That is, making the Hubbard energy 
of the order of or larger than the disorder potential does not necessarily lead to 
complete screening of the disorder. As an
example note that for $U/t=8$ and $W/t=4$ for 3/8 filling, one finds a bimodal distribution
of the local charge densities peaked at roughly 0.5 (for the A sites) and 1.0 (for the B sites).

One important element of work towards the understanding of such systems is the characterization 
of electronic screening of the (impurity) disorder potential, 
and recent publications\cite{dobro,shapiro07} have stressed the importance of disorder screening 
in interacting systems. We have examined the simplest variant of such an approach.
To be specific, for {\em non-interacting} electrons we have determined the ground states 
with screened impurity potentials, {\em viz.} for effective on-site energies given by
\begin{equation}
\label{eq:effVi}
V_i~\rightarrow ~V^{eff}_i~=~\frac{V_i}{\varepsilon}
\end{equation}
Solving for the local charge densities for such a model, we fit the static dielectric
constant, $\varepsilon$, for each filling and $W/t$ by minimizing the mean-squared 
difference of these densities in comparison
to the exact local charge densities. This procedure is quite successful, especially (as expected)
at 1/2 filling, and in TABLE~\ref{table:chargesVeq12halffilling} we compare charge densities for 
$W/t$=12 from which one can see the impressive agreement (note that this is for $U/t=8$) 
between these approaches. For this parameter set and filling, we find $\varepsilon \approx 2.6$.

The simplest and most common approach to finding approximate solutions for interacting
systems is that of using a
self-consistent, real-space unrestricted Hartree-Fock (HF) approach, which previously have been 
employed in various ways in previous studies\cite{imadaRMP,fazilehPRL06,trivedi04}. 

In such a HF approach one ignores terms that are proportional to fluctuations 
about mean values squared, and thus one approximates the local Hubbard interaction being replaced by
(here we ignore the possibility of local superconducting pairing correlations)
\begin{widetext}
\begin{eqnarray}
\label{eq:2siteH}
\langle {\hat n}_{i,\uparrow}{\hat n}_{i,\downarrow} \rangle 
&=& ({\overline n}_{i,\uparrow}
+\delta {\hat n}_{i,\uparrow})({\overline n}_{i,\downarrow} +\delta {\hat n}_{i,\downarrow}) 
- (h^+_i+\delta {\hat h^+}_i) 
(h^-_i+\delta {\hat h^-}_i)\\ \nonumber
&\approx&{\hat n}_{i,\uparrow} {\overline n}_{i,\downarrow} +
{\hat n}_{i,\downarrow} {\overline n}_{i,\uparrow} -
{\overline n}_{i,\uparrow} {\overline n}_{i,\downarrow} -
{\hat h^+}_i h^-_i -
{\hat h^-}_i h^+_i +
h^+_i h^-_i 
\end{eqnarray}
\end{widetext}
where the effective local fields $h_i^\pm$ are given by
\begin{equation}
h_i^+ \equiv \langle {\hat S^+}_i \rangle~~~~~~
h_i^- \equiv \langle {\hat S^-}_i \rangle~~.
\end{equation}
Then, one must find self consistently the local spin-resolved charge densities and local fields
that minimize the variational estimate of the ground state energy. 

\begin{table}
\begin{tabular}{lccr}
(a)      \\[4pt]
1.811 & $\,$ 0.110 & $\,$ 0.219 & $\,$ 1.875 \\[2pt]
0.193 & $\,$ 0.145 & $\,$ 1.821 & $\,$ 1.813 \\[2pt]
0.140 & $\,$ 1.775 & $\,$ 0.143 & $\,$ 0.199 \\[2pt]
1.870 & $\,$ 0.192 & $\,$ 1.819 & $\,$ 1.874 \\[10pt]
(b)      \\[4pt]
1.813 & $\,$ 0.106 & $\,$ 0.231 & $\,$ 1.874 \\[2pt]
0.199 & $\,$ 0.148 & $\,$ 1.816 & $\,$ 1.806 \\[2pt]
0.141 & $\,$ 1.759 & $\,$ 0.147 & $\,$ 0.199 \\[2pt]
1.873 & $\,$ 0.199 & $\,$ 1.811 & $\,$ 1.877 \\
\\
\end{tabular}
\caption{\label{table:chargesVeq12halffilling} In (a) the exact local charge densities are given for
the $U/t=8$ interacting ground state at 1/2 filling for $W/t=12$; (b) shows analogous
data, but now for the exact non-interacting system in which the impurity potential $W/t=12$ is
screened as in Eq.~(\protect\ref{eq:effVi}), for a static, homogeneous dielectric constant
of $\varepsilon = 2.6$.}
\end{table}

\begin{figure}[t]
\centering
\includegraphics[height=6.5cm,width=8.25cm]{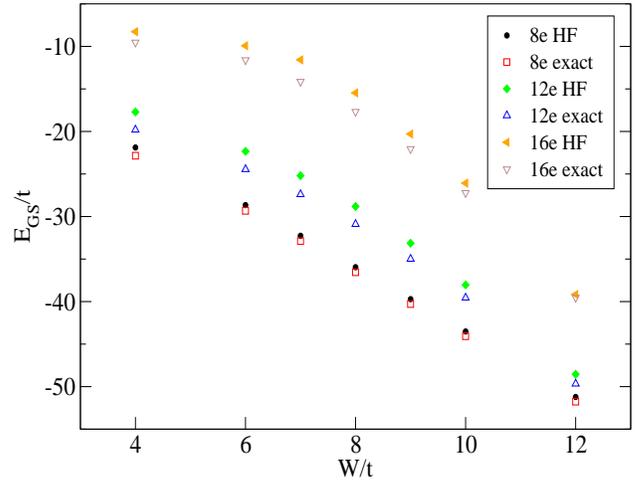}
\caption{\label{fig2:exactvsHFnrgs} [Colour online]
The exact {\em vs.} real-space self-consistent Hartree-Fock energies for the
binary alloy Hamiltonian discussed in the text. The Hubbard interaction
is fixed to be $U/t=8$, and the lower/middle/upper curves correspond to
fixed electronic densities of one-quarter/three-eighths/half filling, namely
8, 12 and 16 electrons (8e, 12e and 16e).
}
\end{figure}

The HF approach is a variational one in which the lowest energy solution within the
product state Hilbert space is found. A comparison of the energies for different
fillings and disorder is shown in Fig.~\ref{fig2:exactvsHFnrgs}, and the agreement
between the exact and HF energies is seen to improve with increasing disorder,
as expected.

It is common rubric that HF works best for disordered systems
(in comparison to ordered systems),  but the critiquing
of such a belief requires that one be in possession of exact solutions with which one can compare
approximate solutions. One such comparison is found in
part (c) of TABLE~\ref{table:chargesVeq4halffilling} shows the local charge densities for
the HF ground state for $W/t=4$ and $U/t=8$ for 1/2 filling (the spins are found to be 
collinear for this state, so it is permissible to set $h_i=0$ -- see below). 
Clearly the agreement between the exact and HF ground
states is very good -- the maximum absolute difference between exact and HF charge 
densities is 0.019. This quantitative agreement is found to be repeated \underline{for all} 
$W/t$, $U/t$ and electronic fillings. One particular example of the success of this approximation
is found for $W/t=12$, $U/t=8$ at 1/4 filling -- for these parameters the largest absolute 
difference between the exact and HF local charge densities is only 0.002. 

We now turn to what one can learn from the HF solutions concerning the spin degrees of
freedom for such systems, both the spin-spin correlations and the spatial arrangement
and orientations of the spins. For the HF solutions
there is some freedom of choice in specifying the possible effective fields that are allowed.
That is, if the $h_i^\pm$ are set to zero then the only nonzero expectation values of 
$\langle {\hat {\bf S}}_i \cdot {\hat e}_\alpha \rangle$, where $\alpha=x,y,z$, are for the 
$z$ components. Such solutions correspond to (potentially) non-paramagnetic spin arrangements
with collinear spins.  If the $h_i^\pm$ are allowed to be non-zero but constrained to be real then $h_i^+ = h_i^-$
and both the $x$ and $z$ components can be nonzero. This allows for the possibility of
the HF solutions having non-collinear (but still coplanar) spins lying in the $xz$ plane -- below we 
will refer to such solutions as ``twisted spin configurations". 
Both of these circumstances lead to restricted 
HF solutions; only when the $h_i^\pm$ are allowed to assume complex values does one have
the fully unrestricted solutions. In our numerical work on
a 4$\times 4$ lattice, we have found that for all system
parameters and fillings the HF states correspond to either collinear 
or coplanar spin configurations.

In contrast to the success of the HF wave functions in producing an excellent
approximation for the local charge densities of the exact ground state, 
since these wave functions are product states
the spin degrees of freedom are essentially classical (although the moments on
each site can be different than 1/2) and therefore the effectiveness of this approximation 
in reproducing the spin correlations in the ground state is not expected to be as good. 
Indeed this is what we find -- there is only a reasonable {\em qualitative} similarity 
between the exact and HF near-neighbour spin-spin correlations. As one typical example note
that for $W/t=6$ and $U/t=8$ at 1/2 filling, the average (absolute) difference between 
the exact and HF evaluations of $\langle {\bf S}_i\cdot {\bf S}_j\rangle$ for $i,j$ being
near neighbours is almost 20\% -- for some near neighbours the HF correlations do not
even have the same sign as found in the exact solutions. Further, comparing the exact and HF
results using a spatially averaged mean-squared difference, while the local charge densities improve with 
increasing disorder, similar to the energies of Fig.~\ref{fig2:exactvsHFnrgs}, the spin-spin
correlations become increasingly worse with increasing disorder. 

The spin degrees of freedom that are present in the HF solutions do contain some potentially important 
information, and here we mention one possible relationship of our results to other studies.
That is, the spin arrangement that we find may have a bearing on a recent proposal for a 
novel metallic phase in two dimensions.\cite{trivedi04} The authors of this paper found that 
at 1/2 filling and for a disorder strength that was approximately equal to the Hubbard energy
there was a significant delocalization of the effective one-electron HF eigenstates at the Fermi 
energy, and thereby forecast the existence of a metallic state in two dimensions. 
An understanding of the physics of this result remains an outstanding problem.

\begin{figure}
\begin{center}
\includegraphics[height=6.5cm,width=6.5cm,clip=true]{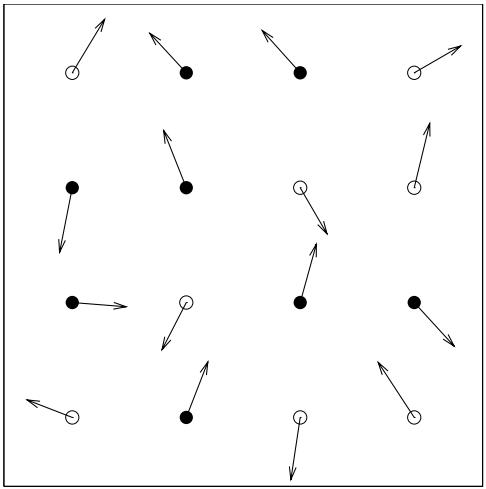}
\caption{\label{fig4:spintwists} 
The spin configuration for a 1/2 filled system with $W/t=7$ and $U/t=8$. The filled circles denote
A sites and the open circles denote B sites. The length of each vector is proportional
to the magnitude of the magnetic moment on that site. 
}
\end{center}
\end{figure}

We have found an interesting result in our HF studies that relates to this problem, that being
that when we focus on 1/2 filling, only for $W/t=$7 and 8 (for $U/t=8$) do we find non-collinear
spin configurations; for all other values of $W/t$ we find collinear HF ground
states. In Fig.~\ref{fig4:spintwists} we show the non-collinear spin configuration found for
1/2 filling for $W/t=7$ and $U/t=8$. It is in this same parameter regime that these 
authors\cite{trivedi04} found the existence of a novel metallic phase, and the role that these 
spin configurations may play in such a transition is an interesting question to explore. 
Further, some evidence of this physics is also found in a comparison of the exact and HF wave functions. 
Out of all of the exact wave functions that we obtained, the only (non-singlet) triplet ground 
state was for 1/2-filling and $W/t=7$. In fact, for this system the HF state had 
a net magnetic moment of 1, consistent with the exact ground state.

To answer the question of what, if anything, does the non-collinear spin configuration
of the HF solution have to do with the enhanced delocalization of electrons found in
Ref. 15, we draw a parallel to the work studying a single hole moving in an antiferromagnet
background (described by the $t-J$ model).
Speculation of the potential physics behind the metallization in two dimensions 
found in Ref. 15 can then be gleaned from the dipolar backflow analysis of a vacancy moving in
an antiferromagnetic background.\cite{dipolar} That is, the spin ``twists" are generated 
by allowing for the mobile vacancy to minimize its energy. Such a state produces a 
non-zero quasiparticle weight, itself a requirement for a metallic phase. 

Summarizing our results, we have completed the exact
diagonalization for a 4$\times$4 cluster with periodic boundary conditions
for a binary alloy with a repulsive on-site Hubbard interaction. 
Our analysis has shown that the interactions generate an effective screening 
of the disorder potential for some fillings and disorder. Further, our real-space
self-consistent Hartree-Fock (HF) solutions are seen to be able to reproduce
quantitatively the local charge densities with considerable success, but not 
the quantitative details of the spin correlations. Our HF solutions display
non-collinear, coplanar spin twists in the ground state at 1/2 filling when 
the disorder and interaction energies are comparable. 
This report will be followed by a longer paper with
many more numerical results and detailed analysis.

We thank Avid Farhoodfar, Bill Atkinson and Nandini Trivedi for helpful discussions.
This work was supported in part by the Hong Kong RGC and the NSERC of Canada.

\end{document}